\documentclass[prl,twocolumn,showpacs,preprintnumbers,amsmath,amssymb]{revtex4}

\usepackage{color}
\usepackage{graphicx}
\usepackage{dcolumn}
\usepackage{bm}

\preprint{submitted to Journal of Spintronics and Magnetic
Nanomaterials}

\begin{document}

\title{Suppression of magnetism in BiFeO$_3$ ultrathin epitaxial multilayers}

\author{K. Koumpouras}
\author{I. Galanakis}\email{galanakis@upatras.gr}

 \affiliation{Department of Materials Science, School of Natural Sciences,
University of Patras,  GR-26504 Patra, Greece}

\date{\today}

\begin{abstract}
Preliminary first-principles calculations on the magnetic behavior
of ultra-thin epitaxial multilayers between the BiFeO$_3$
magnetoelectric compound and various types of spacers are
presented. As spacer we have considered i) InP semiconductor, ii)
Fe which is a ferromagnet, and iii) metallic V. In all cases under
study the growth axis of the multilayer was the [001]. Our results
indicate that the magnetic properties are seriously downgraded for
the ultrathin BiFeO$_3$ multilayers independent of the nature of
the spacer and in some cases under study magnetism even vanishes.
More extensive calculations are needed to establish a more clear
view of the physical properties of the interfaces involving the
BiFeO$_3$ compound. The present manuscript completes the study
presented in two recent research articles [K. Koumpouras and I.
Galanakis, \textit{J. Magn. Magn. Mater.} 323, 2328 (2011);
\textit{ibid}, \textit{J. Spintron. Magn. Nanomater.} 1, in press
(2012)].
\end{abstract}


\maketitle
\section{Introduction}

 Among the most interesting new classes of materials under intense
investigation in Spintronics\cite{Zutic,Felser,Zabel} are the
so-called multiferroics which combine several ferroic orders like
ferromagnetism, ferroelectricity, ferroelasticity
etc.\cite{Review,Review5,Review6} The compounds combining the
electric and magnetic order\cite{Review}  have several potential
applications like magnetic-field sensors and electric-write
magnetic-read random-access
memories.\cite{Hetero,General1,Zhang,Review7,Review8} Magnetic
order and ferroelectricity have different
origins\cite{Picozzi,Picozzi2,General2} and thus the materials
exhibiting the magnetoelectric effect are few and the coupling
between the magnetic and electric properties is weak. An
alternative route to achieve a strong coupling could be the growth
of thin film heterostructures and several advances have been made
towards the magnetic control of
ferroelectricity\cite{General3,General4,Review2} and the electric
control of thin film magnetism.\cite{Films3,Review3,Review4}

Bismuth ferrite is probably the most studied representative of
magnetoelectric materials. Bulk BiFeO$_3$ crystallizes in a
perovskite-like pseudocubic structure instead of a ferrite one and
is classified as a ferroelectric G-type
antiferromagnet.\cite{Kiselev}   Several first-principles
calculations have been carried out to study the properties of bulk
BiFeO$_3$\cite{BiFeO3-Calc} and we refer readers to Ref.
\onlinecite{Palova} for an overview of the literature on this
compounds. Since the single-component crystals like BiFeO$_3$
present only a weak magnetoelectric effect, an alternative route
to achieve a more strong effect has been proposed to be  the
growth of heterostructures where epitaxial strain can enhance the
phenomenon.\cite{Films1,Trilayers,Kovachev,Yamauchi} Thus
first-principles calculations of such heterostructures involving
alternating layers of BiFeO$_3$ and various spacers can serve as a
test-ground to study the behavior of electric and magnetic
properties of films. The study of the latter in the case of
ultrathin epitaxial films is the aim of the present manuscript.

In a recent publication (Ref. \onlinecite{Koumpouras}) we have
presented extended first-principles calculations, employing the
Quantum-ESPRESSO\cite{QE} ab-initio electronic structure method in
conjunction with the Generalized-Gradient Approximation (GGA) in
the Perdew-Burke-Erzenhof formulation,\cite{GGA} on the electronic
and magnetic properties of BiFeO$_3$ alloy as a function of the
lattice constant in the case of the cubic perovskite structure
(see figure 1 in Ref. \onlinecite{Koumpouras}). In Ref.
\onlinecite{Koumpouras2} we have extended this study to cover also
the case of Mn substation for Fe. All compounds under study in
these two references exhibited significant magnetic properties
with spin magnetic moments at the Fe and Mn sites of the order of
3 $\mu_B$. In the present contribution we expand these two studies
using the same ab-initio method with grids of the same density in
the \textbf{k}-space to cover also the case of ultrathin epitaxial
films. Although these films are ideal cases and cannot be, with
few exceptions, be realized experimentally, they can serve as
starting point to understand the interplay between the interface
structure and the magnetic properties of BiFeO$_3$ films with
other spacers. For completeness we took into account three
different spacers to cover different cases of electronic
properties at the interface: InP semiconductor, metallic V and
ferromagnetic Fe. In Section II we present the structure of the
multilayers and the results for the case of the InP spacer, in
section III the cases of Fe and V metallic spacers and finally in
Section IV we present the main conclusions of our work.

 \section{I\lowercase{n}P spacer}
\subsection{Structure of the multilayer}

\begin{figure}
\includegraphics[width=\columnwidth]{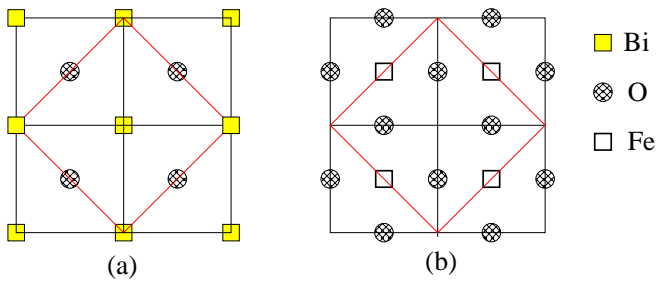}
\caption{Schematic representation of the structure of the BiO (a)
and FeO$_2$ (b) possible terminations of the BiFeO$_3$ alloy along
the [001] direction. With the red lines we denote the limits of
the two-dimensional unit cell of InP (see Fig. \ref{fig2}) which
has a lattice constant: $a_{InP}=\sqrt{2}\times a_{BiFeO_3}$. }
\label{fig1}
\end{figure}

\begin{figure}
\includegraphics[width=\columnwidth]{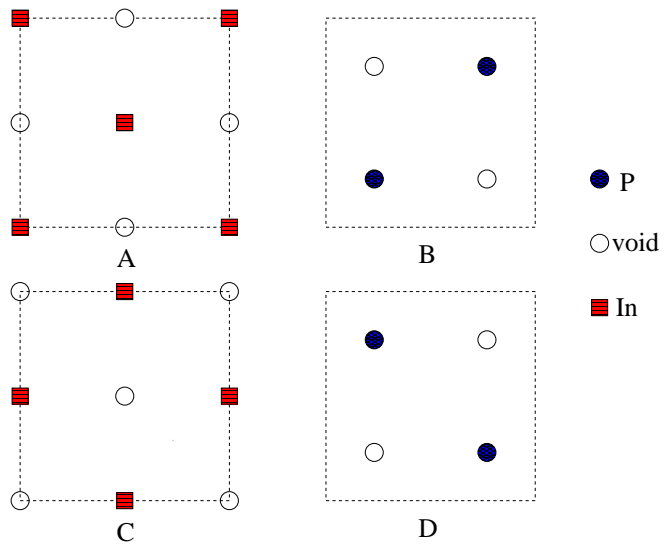}
\caption{Structure of the atomic layers of InP along the [001]
direction taking into account also the vacant sites. The structure
is built using four alternating pure atomic layers A, B, C and D.
The difference between A(B) and C(D) is that the In(P) atoms and
the voids have exchanged sites.} \label{fig2}
\end{figure}

Along the [001] direction BiFeO$_3$  can be considered to consist
of alternating layers of BiO and FeO$_2$ as shown in Fig.
\ref{fig1}.    The in-plane unit cell is a square of the same
lattice constant as the cubic unit cell of the bulk BiFeO$_3$. InP
crystallizes in the zincblende    structure, similar to GaAs, as
most of the III-V semiconductors, and has an experimental lattice
constants of 5.87 \AA . Along the [001] direction, the zincblende
structure can be viewed as consisting of alternating pure In and
pure P atomic layers as shown in Fig. \ref{fig2}. To fully
describe the      zincblende structure we have to consider, except
the In and P atoms, also the occurrence of vacant sites (voids);
there are exactly two no-equivalent  voids within the zincblende
unit cell. As shown in Fig. \ref{fig2} along the [001] direction
there are not two but four alternating non-equivalent atomic
layers. The difference between the A(B) and C(D) layers is that
the In(P) atoms have exchanged sites with the voids. This exchange
of atomic       positions is important since it lead to four
non-equivalent interfaces between InP and BiFeO$_3$ in our study.
Finally we have to mention that the lattice constant of InP
coincides with the BiFeO$_3$ lattice constant multiplied by the
square root of two: $a_{InP}=\sqrt{2} \times a_{BiFeO_3}$. Thus as
shown in Fig. \ref{fig1} the diagonal of the BiFeO$_3$
two-dimensional square unit cell can be assumed to be the side of
the corresponding unit cell of the InP (denoted with red color in
the figure) and thus epitaxial growth between InP and BiFeO$_3$
can be assumed.

For the  InP/BiFeO$_3$ multilayers we studied four different cases
with respect to the relative position of the atoms at the
interface between the two spacers. Along the growth axis we took
into account eight atomics layers which are repeated in the [001]
direction. Initially we studied the structure
...FeO$_2$//In/P/In/P/BiO/FeO$_2$/BiO/FeO$_2$//In..., and thus in
the first case we have two non-equivalent interface in our
structure: a P/Bi and an In/Fe ones. A close examination of Figs.
\ref{fig1} and \ref{fig2} reveals that, in the case just mentioned
:(a) with respect to the P/Bi interface P atoms are situated in
the diagonal connecting Bi atoms (compare (a) in Fig. \ref{fig1}
with the D layer in Fig. \ref{fig2}), and (2) with respect to the
In/Fe interface In atoms are situated in the midpoint between Fe
nearest neighbors (take into account A layer in Fig. \ref{fig2}
and combine it with the (b) in Fig. \ref{fig1}).

In the second case under study In and P atoms have exchanged sites
with respect to the first case under study and thus now we have
along the [001] direction a
...FeO$_2$//P/In/P/In/BiO/FeO$_2$/BiO/FeO$_2$//P... structure,
\textit{i.e.} the inequivalent interfaces are now In/Bi and Fe/P.
Notice that as in the Fe/In interface in case 1,  also in the P/Fe
interface, which occurs in case 2, the P atoms at the interface
are located in the midpoints between the Fe atoms (if we examine
only the in-plane projection of the multilayer) just above the
oxygen atoms and this results to vanishing magnetism in case 2
under study as will be discussed later in the next subsection.

As we mentioned above the structure of InP along the [001]
direction should be viewed as consisting of four atomic layers
(see Fig. \ref{fig2}) since voids play a crucial role in
interfaces. If in the two previous cases 1 and 2 we exchange layer
A(B) with layer C(D) in Fig. \ref{fig2} we get two new cases 3 and
4 with different arrangement of the atoms at the interface. In
case 3(4), the succession of the atomic layers is similar to case
1(2). When we compare case 1(2) with case 3(4), we conclude that
the P(In)/Bi interfaces are similar since the P(In) atoms are now
situated at the other diagonal connecting the Bi atoms. Critical
is  the other In(P)/FeO$_2$ interface since now in case 3(4) the
In(P) atoms are located just above the Fe atoms while in case 1(2)
they were located above the oxygen atoms at the midpoints between
neighboring Fe atoms.

\subsection{Results and discussion}

In all cases under study our results converged to the
ferromagnetic solution independently of the initial conditions and
initial arrangement of the spin moments. As in-plane lattice
constant we have chosen the $\sqrt(2) \times 4.153$ \AA\ $=5.874$
\AA\ (11.1 a.u.) which is the experimental lattice constant of InP
and moreover for the lattice constant of 4.153 \AA\ BiFeO$_3$
exhibits very pronounced magnetic properties.\cite{Koumpouras} As
a result of the epitaxial growth along the [001] axis the lattice
constant was 14.182 \AA\ (26.8 a.u.).

\begin{figure}
\includegraphics[width=\columnwidth]{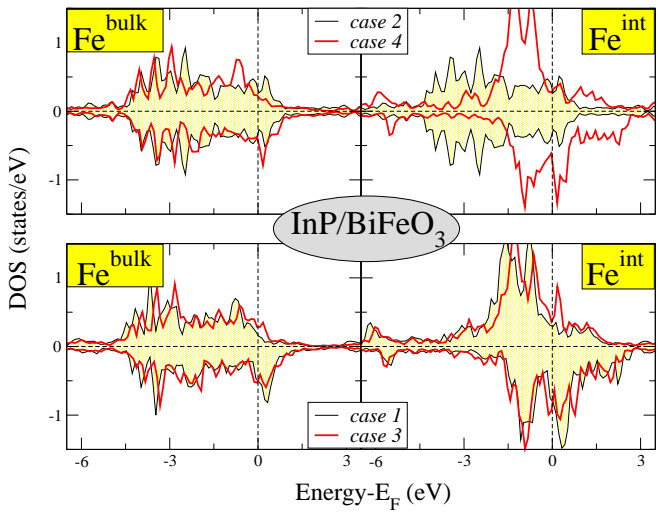}
\caption{Density of states (DOS) projected on the Fe d-orbitals
for the
 Fe atoms in the BiFeO$_3$ spacer (Fe$^{bulk}$ at the middle of the spacer
 and Fe$^{int}$ at the FeO$_2$ interface for all four cases under study (see text
 for explanation). Positive DOS corresponds to the majority spin (spin-up) electrons
 and negative DOS to the minority spin (spin-down) electrons. The Fermi level has been
 assigned to the zero energy.  } \label{fig3}
\end{figure}

In Fig. \ref{fig3} we present the density of states (DOS) of the
Fe d-orbitals  for all four cases under study. We denote as
Fe$^{bulk}$ the Fe atoms within the BiFeO$_3$ spacer and with
Fe$^{int}$ the Fe atoms located at the interface. We have also to
note here that for the multilayer structures under study, we have
a complete lift of the degeneracy of the Fe d-orbitals and we
cannot refer anymore to the double-degenerated $e_g$ and triple
degenerated $t_{2g}$ orbitals as in bulk BiFeO$_3$.
 In case 2 where we have a P/Bi interface our calculations have
converged to a non-magnetic solution. The loss of magnetism in
this case should be attributed to the reduced hybridization
between the neighboring Fe and P atoms at the interface. The
latter have as valence 3p electrons which are less extended in
space with respect to the In 5p valence states and do not
hybridize with the Fe d-orbitals at the interface. In case I, the
d-band is shifted to higher energy values for the spin-down
electrons with respect to the spin-up electrons. This is more easy
to visualize for the Fe$^{int}$ atoms since their  hybridization
with the p-orbitals of In leads to more pronounced electronic
properties (smaller bandwidth with respect to the Fe$^{bulk}$
atoms and thus more intense picks). The latter is also reflected
on the spin magnetic moments presented in Table \ref{table1} where
the spin magnetic moments are considerable larger for Fe$^{int}$
with respect to Fe$^{bulk}$.

\begin{table}
\caption{Fe spin magnetic moments in $\mu_B$ within the BiFeO$_3$
spacer for the InP/BiFeO$_3$ epitaxial ultrathin multilayers for
all four cases under study (see text for explanation of different
cases). The BiFeO$_3$ spacer contains two inequivalent atoms:
Fe$^{bulk}$ at the middle of the BiFeO$_3$ spacer and Fe$^{int}$
at the interface.} \label{table1}
\begin{ruledtabular}
 \begin{tabular}{lcccccccccccc}

  &  Fe$^{bulk}$    &  Fe$^{int}$ \\
  Case 1 &0.357 & 1.158 \\
  Case 2 & 0 & 0 \\
  Case 3 & 0.247& 0.746 \\
  Case 4 & 0.296& 0.919

\end{tabular}
\end{ruledtabular}
\end{table}

In Fig. \ref{fig3} we also present the DOS of the Fe atoms for the
cases 3 and 4 where we have changed the positions of the In(P)
atoms at the interface. More precisely, as we discussed  in the
previous section, the In(P) atoms at the interface with FeO are
not any more situated above the Fe atoms instead of the O atoms in
cases 1 and 2. This leads to increased hybridization also in case
4, where we have a P/Fe interface and now the multilayer converges
to a magnetic solution contrary to case 2 where we had a
non-magnetic configuration. If we compare cases 1 and 3, where the
Fe/In contact appears, we can notice that in case 3 the splitting
of the Fe$^{int}$ d-bands is smaller due to the alteration in the
Fe 3d-In 5p orbitals hybridization resulting also in smaller spin
magnetic moments as shown in Table \ref{table1}. But overall the
obtained DOS are similar to the ones obtained for case 1 and
presented also in Fig. \ref{fig3}.

\section{V and F\lowercase{e} spacers}

\begin{figure}
\includegraphics[width=\columnwidth]{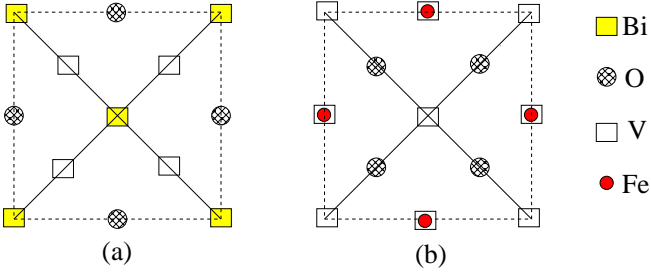}
\caption{Schematic representations of the V/BiO (a) and V/FeO$_2$
interfaces in the case of the V/BiFeO$_3$ multilayer. For the
Fe/BiFeO$_3$ multilayer under study Fe atoms simply substitute the
V atoms since both V and Fe have a bcc structure. } \label{fig4}
\end{figure}

We continue our study using two metallic spacers, ferromagnetic Fe
and normal metallic V, instead of InP. Both Fe and V crystallize
in the bcc structure. The zincblende structure of InP, if we take
into account the vacant sites and ignore the different chemical
species, is in reality a bcc one. Thus the structure of the
Fe/BiFeO$_3$ and V/BiFeO$_3$ multilayers is similar to the
structure of the InP/BiFeO$_3$ where all sites are occupied
exclusively by Fe or V atoms.  We have studied 5 cases for both
multilayers where we have just varied the lattice constant. Case 1
corresponds to the same lattice constant as for InP/BiFeO$_3$
studied in the previous section. Cases 2 and 3 correspond to a
uniform compression of the lattice parameter in all directions of
10 and 5 \% , respectively, and cases 4 and 5 of a uniform
expansion of the lattice parameter of 5 and 10 \% , respectively.
Our aim is to     study the behavior of the magnetic properties
upon hyrdrostatic pressure. In cases 1, 4 and 5 we converged to a
ferromagnetic solution irrespectively of the initial arrangement
of the spin magnetic moments. In contrast in the compressed cases
2 and 3 for Fe/BiFeO$_3$  we could not get convergence while for
the case of the V spacer we converged to a non magnetic solution.
This behavior stems from the reduced volume around the Fe atoms
which leaded to a large compression of the Fe d-orbitals and
suppression of magnetism as expected by the magnetovolume effect
which has been extensively studied for transition metal atoms. In
the multilayers under study we have also two inequivalent
interfaces: a V(Fe)/BiO and a V(Fe)/FeO$_2$ contact. In the first
interface the four V(Fe) atoms are located within the two
diagonals connecting the Bi atoms as shown in Fig. \ref{fig4}a
while in the second interface they are located just above the Fe
atoms of FeO$_2$ as shown in Fig. \ref{fig4}b.

\begin{figure}
\includegraphics[width=\columnwidth]{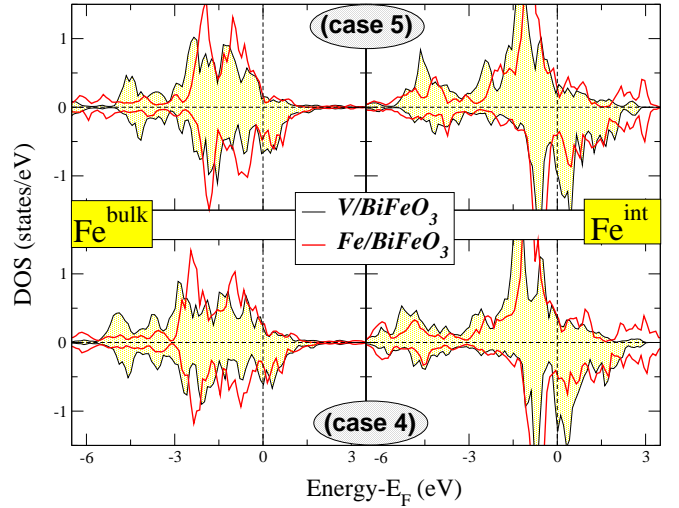}
\caption{DOS projected on the Fe d-orbitals for the Fe atoms
within the BiFeO$_3$ spacer for both V/BiFeO$_3$ and Fe/BiFeO$_3$
multilayers. Results are for the two expanded lattice parameters
denoted as cases 4 and 5 (see text for details). Details as in
Fig. \ref{fig3}. } \label{fig5}
\end{figure}

We will start our discussion concentrating on the Fe atoms within
the BiFeO$_3$ spacer in the case of the Fe/BiFeO$_3$ multilayer
and in Table \ref{table2} we have gathered the Fe spin magnetic
moments for both Fe atoms in the inside of the BiFeO$_3$ spacer
(Fe$^{bulk}$) and at the interface (Fe$^{int}$).  As mentioned
above for the compressed cases 2 and 3 we were not able to
converge to a solution. For case 1 we converged to a non-magnetic
solution and for the more expanded lattice parameters (cases 4 and
5) we converged to a magnetic solution. The spin magnetic moments
are larger for the more expanded case 5. This behavior is expected
by the well studied     magnetovolume effect. For the late
transition metal atoms like Fe, the spin-splitting of the d-states
increases with the atomic volume. Thus as we expand the lattice
the tendency to magnetism increases leading to larger spin moments
while compression of the lattice eventually leads to loss of
magnetism.  A similar situation occurs also for the case of
V/BiFeO$_3$ but now the tendency to magnetism is stronger for the
same lattice parameter.        For cases 2 and 3 we converged to a
non-magnetic solution while we got a ferromagnetic configuration
for case 1 contrary to the Fe/BiFeO$_3$ multilayer. Fe spin
magnetic moments are larger for the V spacer but the largest
calculated value, which we got for Fe$^{int}$ in case 5 as shown
in Table \ref{table2}, is 1.3 $\mu_B$ almost half the value in
pure bulk Fe. Although magnetism is present for the cases with the
more expanded lattice           constant the magnetic properties
are seriously downgraded with respect to pure BiFeO$_3$ bulk
crystals.

\begin{table}
\caption{Fe spin magnetic moments in $\mu_B$ within the BiFeO$_3$
spacer for the V(Fe)/BiFeO$_3$ epitaxial ultrathin multilayers for
all cases under study. Case 1 corresponds to a BiFeO$_3$ in-plane
lattice constant of 4.153 \AA , cases 2 and 3 to uniform
compression by 10 and 5 \%\ respectively, and cases 4 and 5
uniform expansion by 5 and 10 \%\ respectively. Zero values means
that we have converged to a non-magnetic solution and "--" that we
were not able to achieve convergence. Details as in Table
\ref{table1}.} \label{table2}
\begin{ruledtabular}
 \begin{tabular}{lcccccccccccc}
& \multicolumn{2}{c}{V/BiFeO$_3$} & \multicolumn{2}{c}{VFe/BiFeO$_3$} \\
  &  Fe$^{Bulk}$    &  Fe$^{int}$ &  Fe$^{Bulk}$    &  Fe$^{int}$\\
  Case 1 & 0.146 & 0.777 & 0 & 0 \\
  Case 2 & 0 & 0 & -- & -- \\
  Case 3 & 0 & 0 & -- & -- \\
  Case 4 & 0.269 & 1.252 & 0.224 & 0.620 \\
  Case 5 & 0.325 & 1.301 & 0.339 & 0.815

\end{tabular}
\end{ruledtabular}
\end{table}

In the case 1 of Fe/BiFeO$_3$, presented in the lower panel of
Fig. \ref{fig7}, both Fe atoms within the BiFeO$_3$ spacer are non
magnetic and the DOS is the same for both spin directions, while
in the case of V/BiFeO$_3$ a small splitting of the d-bands
appears in accordance to the spin magnetic moments presented in
Table \ref{table2}. This splitting increases as we move to case 4
and 5 (presented in Fig. \ref{fig5}) following the increase of the
spin magnetic moments. We can make two remarks with respect to the
presented DOS. First, for the Fe atoms at the interface
(Fe$^{int}$) the weight of the d-states is shifted close to the
Fermi level with respect to the Fe$^{bulk}$ atoms as a consequence
of the increased hybridization at the interface. Second, if we
compare the behavior of the same Fe atom for the same lattice
constant in the two multilayers under study, the exchange
splitting of the d-bands is larger in the case of the V spacer
(spin-up states are deeper in energy and spin-down states higher
in energy) in accordance with the larger spin magnetic moments in
this case.

\subsection{Effect of the change of V(Fe) positions at the interface}

\begin{figure}
\includegraphics[width=\columnwidth]{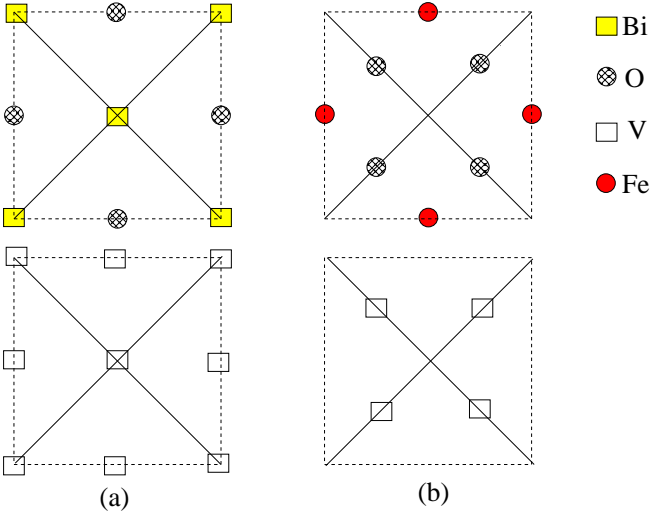}
\caption{Same as Fig. \ref{fig4} but now we have exchanged the V
atomic layers. In the left (a) panel the V/BiO interface and in
the right panel (b) the V/FeO$_2$ interface. Notice that with
respect to Fig. \ref{fig4}, the V atoms are located now exactly at
the top of the O and Bi atoms. We do not present the V layers and
the BiO(FeO$_2$) layers in the same square as in Fig. \ref{fig4}
to make the atomic positions more clear.} \label{fig6}
\end{figure}

\begin{figure}
\includegraphics[width=\columnwidth]{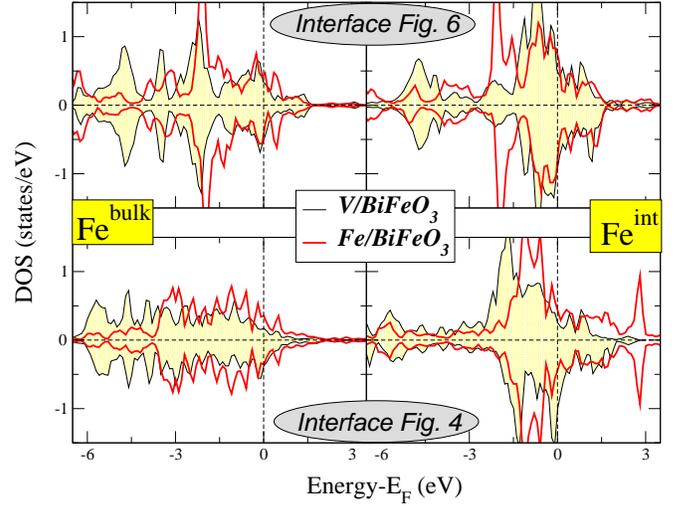}
\caption{DOS projected on the Fe d-orbitals for the Fe atoms
within the BiFeO$_3$ spacer for both V/BiFeO$_3$ and Fe/BiFeO$_3$
multilayers. Results are for the lattice constant of 4.153 \AA .
In the bottom panel we present the results for the interface
structure of Fig. \ref{fig4} and in the upper panel for the
interface structure in Fig. \ref{fig6}. Details as in Fig.
\ref{fig3}. } \label{fig7}
\end{figure}

In this subsection we present results again for the V/BiFeO$_3$
and Fe/BiFeO$_3$ multilayers for the lattice constant of case 5
for which magnetism is more pronounced. And thus the in-plane
lattice constant is 6.46 \AA\ and  the out of plane 15.61 \AA\ .
The difference with the multilayers studied just above is that we
have changed the position of the V(Fe) atoms at the interface. As
shown in Fig. \ref{fig6} the V(Fe) atom at the interface with FeO
are now located just above the  oxygen atoms of the BiFeO$_3$
spacer and not the Fe ones. The change of the local environment
leads to reduced hybridization of d-orbitals of the
transition-metal atoms at the interface due  to the larger
Fe-V(Fe) distance and thus to smaller induced spin magnetic
moments in the BiFeO$_3$ layer. For both V/ and Fe/BiFeO$_3$
multilayers the Fe$^{int}$ spin magnetic moment is 0.26 $\mu_B$,
while the Fe$^{bulk}$ moment  is 0.11 $\mu_B$ for the case of the
V spacer and only 0.002 $mu_B$ for the case of the Fe spacer. Our
discussion on the spin magnetic moments is reflected also on the
DOS presented in the upper panel of Fig. \ref{fig7} where the
imbalance between the spin-up and spin-down states is very small.
The only noticeable effect is the shift of the Fe$^{int}$ d-states
between the two multilayers; for the V spacer the Fe$^{int}$
states are more concentrated around the Fermi level. The
Fe$^{bulk}$ atoms are shielded from the interface due to the
surrounding Bi and O atoms and the DOS is more similar for both
type of V and Fe spacers as reflected also on
 the spin magnetic moments.

\subsection{Behavior of V and F\lowercase{e} atoms within the
V(F\lowercase{e})spacer}

In this section we will shortly refer to the magnetic properties
of the V and Fe atoms within the transition metal spacers. There
are four transition metal atoms within each atomic layer and we
have 4 atomic layers in our spacer which we count starting from
the interface with the FeO$_2$ (we denote it as layer 1) and end
with the atomic layer at the interface with BiO.  We will start
our discussion for the V spacer in the case of the V/BiFeO$_3$
multilayer. V atoms present vanishing spin magnetic moments as
shown in Table \ref{table3} in all cases under study and only in
the case 5 (largest lattice constant under study), do the V atoms
in the layer close to the FeO$_2$ layer show a small spin magnetic
moment of about 0.13 $\mu_B$.

\begin{table}
\caption{V(Fe) spin magnetic moments in $\mu_B$ within the V(Fe)
spacer for the V(Fe)/BiFeO$_3$ epitaxial ultrathin multilayers. We
do not present results for the compressed cases 2 and 3 since the
V/BiFeO$_3$ was found non-magnetic while for Fe/BiFeO$_3$ we could
not get convergence. We have 4 atomic layers of V(Fe) spacer and
each one contains four V(Fe) atoms as shown in Fig. \ref{fig4}. We
count the layers starting from the one at the V(Fe)/FeO$_2$
interface and finishing at the V(Fe)/BiO interface. With "case
5-II" we denote the structure presented in Fig. \ref{fig6}. Notice
that within each layer there are two inequivalent with respect to
their magnetic properties Fe(V) atoms, \textit{e.g} at the
V/FeO$_2$ interface presented in Fig. \ref{fig4} there are the V
atoms at the diagonals located just above at the Fe atoms and the
V atoms at the corners and the middle of the square unit cell.
Near the BiO interface all V(Fe) atoms as shown in Fig. \ref{fig4}
have the same nearest neighbors environment and for all atoms in
layer 4 close to the V(Fe)/BiO interface we got the same value of
the spin magnetic moment for all V(Fe) atoms within the layer.}
\label{table3}
\begin{ruledtabular}
 \begin{tabular}{lrrrrrrrccccccc}
& \multicolumn{4}{c}{V/BiFeO$_3$}  \\
  &  case 1       &  case 4 & case 5 & case 5-II\\
  Layer 1 & 0.016   & -0.010& -0.003 & -0.003\\
    & 0.062 &   0.130 & 0.095 & 0.039\\
  Layer 2 & -0.005 &   0.015 & 0.016& 0.007\\
   Layer 3 &  -0.015 & -0.003 & -0.004 & 0.015 \\
          & 0.002  &  -0.006 & -0.009 & -0.002\\
  Layer 4 &  -0.005 &  0.005& 0.010 & 0.003\\
& \multicolumn{4}{c}{Fe/BiFeO$_3$}  \\
  &  case 1       &  case 4 & case 5 & case 5-II \\
  Layer 1 & 0 & -0.026 & -0.129& -0.110\\
    &  &  -0.040 & -0.174 & 0.479\\
  Layer 2 & 0 & 0.017 & 0.209 & -0.023\\
   Layer 3 &  0 &  -0.005 & -0.021 & -0.022 \\
          &  &  -0.012 & -0.202& 0.058\\
  Layer 4 & 0 & 0.015& 0.116 & -0.010\\

\end{tabular}
\end{ruledtabular}
\end{table}

Similar is the situation for the Fe spacer where for the two
smaller lattice constants we could not even converge our
calculations. As shown in Table \ref{table3} only for the largest
value of the lattice constant we got significant values of the Fe
spin magnetic moments, which even for this case, are considerable
smaller than the spin magnetic moments of Fe atoms of the
BiFeO$_3$ spacer ($\sim$0.2 $\mu_B$ for Fe spacer compared to
$\sim$ 0.8 $\mu_B$ for BiFeO$_3$ spacer) and are about one order
of magnitude smaller than in bulk Fe.

We have also included in Table \ref{table3} the results for the
interface structure of Fig. \ref{fig6}, denoted as "case 5-II"
where we have changed the positions of the V(Fe) atoms at the
interface.  Overall also in this case the magnetic properties of
the spacer are not significant. Only in the case of the
Fe/BiFeO$_3$ multilayer half the Fe atoms at layer 1 (located at
the interface with FeO$_2$) present a significant spin magnetic
moment of about 0.5 $\mu_B$ which is still much lower than the
bulk Fe value.

\section{Conclusions}

We expand our study on the magnetic properties BiFeO$_3$ presented
in Refs. \onlinecite{Koumpouras} and \onlinecite{Koumpouras2} to
the case of ultrathin epitaxial multilayers using the
Quantum-ESPRESSO first-principles electronic structure
method.\cite{QE} We have studied several cases of these ultrathin
epitaxial BiFeO$_3$ multilayers using different types of spacers
covering a wide range of electronic materials: InP semiconductor,
ferromagnetic Fe and metallic V. Irrespectively of the spacer the
magnetic properties of the multilayers were found to seriously
degrade due to the low-symmetry of the film and in some cases
magnetism vanished completely. Thus ultrathin multilayers of
BiFeO$_3$ are not suitable for spintronic applications. More
extended calculations covering more thick multilayers and
including quantum molecular dynamics are needed in the future to
determine in more detail the properties of the interfaces of
BiFeO$_3$ with other alloys.

\begin{acknowledgements}
Financial support from the University of Patras (K. Karatehodori
2008 program Nr. C588) is  acknowledged.
\end{acknowledgements}

\end{document}